\documentclass{article}
\usepackage{graphicx}

\begin{document}

\def\giorno{Revised version -- 05/03/2006}

\def\A{{\mathcal A}}
\def\B{{\mathcal B}}
\def\C{{\mathcal C}}
\def\D{{\mathcal D}}
\def\G{{\mathcal G}}
\def\Fb{{\bf F}}
\def\F{{\mathcal F}}
\def\K{{\mathcal K}}
\def\H{{\mathcal H}}
\def\L{{\mathcal L}}
\def\M{{\mathcal M}}
\def\N{{\mathcal N}}
\def\P{{\mathcal P}}
\def\Q{{\mathcal Q}}
\def\R{{\bf R}}
\def\Z{{\bf Z}}
\def\a{\alpha}
\def\b{\beta}
\def\eps{\varepsilon}
\def\de{\delta}
\def\om{\omega}
\def\Om{\Omega}
\def\la{\lambda}
\def\La{\Lambda}
\def\sse{\subseteq}
\def\ss{\subset}
\def\s{\sigma}
\def\S{\Sigma}
\def\ga{\gamma}
\def\Ga{\Gamma}
\def\toro{{\bf T}}
\def\T{{\rm T}}
\def\de{\delta}
\def\d{{\rm d}}
\def\phi{\varphi}
\def\vth{\vartheta}
\def\z{\zeta}
\def\pa{\partial}
\def\w{\wedge}
\def\({\left(}
\def\){\right)}
\def\[{\left[}
\def\]{\right]}
\def\^#1{{\widehat #1}}
\def\~#1{{\widetilde #1}}

\def\Hol{{\mathcal H}}
\def\hol{h}

\def\Remark#1{\medskip \par\noindent {\bf Remark {#1}.}}

\def\sp{\medskip \par\noindent}
\def\EOP{ \hfill $\diamondsuit$ \par \medskip}
\def\EOR{ \hfill $\odot$ \par \medskip}

\def\interno{\hskip 2pt \vbox{\hbox{\vbox to .18
truecm{\vfill\hbox to .25 truecm
{\hfill\hfill}\vfill}\vrule}\hrule}\hskip 2 pt}

\def\mapright#1{\smash{\mathop{\longrightarrow}\limits^{#1}}}
\def\mapdown#1{\Big\downarrow\rlap{$\vcenter{\hbox{$\scriptstyle#1$}}$}}
\def\mapleft#1{\smash{\mathop{\longleftarrow}\limits^{#1}}}
\def\mapup#1{\Big\uparrow\rlap{$\vcenter{\hbox{$\scriptstyle#1$}}$}}


\title{The Poincar\'e--Lyapounov--Nekhoroshev theorem for
involutory systems of vector fields}

\author{Giuseppe Gaeta\\
{\it Dipartimento di Matematica, Universit\`a di Milano} \\
{\it via Saldini 50, I--20133 Milano (Italy)} \\
{\tt gaeta@mat.unimi.it}}

\date{\giorno}
\maketitle

{\bf Abstract} We extend the Poincar\'e-Lyapounov-Nekhoroshev
theorem from torus actions and invariant tori to general
(non-abelian) involutory systems of vector fields and general
invariant manifolds.

\section*{Introduction.}

The celebrated Poincar\'e-Lyapounov theorem gives conditions
ensuring that a periodic solution of a smooth dynamical system is
persistent under small perturbations.

The theorem was extended by Nekhoroshev \cite{Nek} to the case of
quasi-periodic solutions of partially integrable hamiltonian
systems. His result is referred to as the
Poincar\'e-Lyapounov-Nekhoroshev (PLN) theorem.

Detailed proofs of Nekhoroshev's result were provided in
\cite{BamGae} from an analytical point of view, and in
\cite{GaeAOP} from a geometrical one (the latter work also
contains an extension to non-hamiltonian vector fields). The
non-hamiltonian frame was fully considered in \cite{GaeJNMP},
where generalization of results concerning bifurcation from
periodic solutions to bifurcation from quasiperiodic solutions are
also considered.

Further developments include extension to infinite dimensional
systems (and existence of breathers) \cite{BamVel}, to
perturbation of systems with non-compact invariant manifolds
\cite{Fio,FGS}, and to partially integrable bi-hamiltonian systems
\cite{GMS}.

All these results are based on the assumption that the algebra
$\G$ of vector fields under considerations, hamiltonian or
otherwise, is {\it abelian}; and correspondingly the invariant
manifold $\La$ is a torus $\toro^k$ or (in the non-compact case
\cite{GMS}) the product of a torus by a contractible manifold.

After the publication of \cite{GaeAOP} prof. Duistermaat remarked
in a kind letter that, by some arguments based on the geometry of
foliations, one should expect an equivalent result to hold also in
the non-abelian case.\footnote{In the appendix of \cite{GaeAOP} it
was remarked that several parts of the proof of the PLN theorem
given there do not extend to the case where $\G$ is non abelian
and $\La$ is not a torus. The result we obtain here is indeed
weaker than the one holding in the abelian case, and the proof
requires some modification of the arguments used there.} The
purpose of this note is precisely to extend the PLN theorem to the
case of non-abelian algebras of vector fields, and more generally
to involutory systems of vector fields, albeit in a slightly
different way.

Some words are maybe in order, before going into mathematical
detail, about the physical motivation for such a study and
possible physical applications of its results.

The main field of applications for the results obtained here would
be that of (differentiable) dynamical systems -- i.e. systems of
first order ODEs on a smooth manifold. In this framework, one of
the considered vector fields would be the dynamical one, while the
other ones would be symmetries of the former\footnote{Note that
considering these on the same footing corresponds to what is done
in the hamiltonian case, where the dynamical Hamiltonian and those
describing the commuting integrals of motion are treated on equal
basis.}. It should be recalled, indeed, that for first order
(systems of) ODEs, contrary to all other cases, the natural
algebraic structure for the set ${\mathcal G}$ of vector fields
being Lie symmetries of this is not that of a Lie algebra
(${\mathcal G}$ being infinite dimensional as such \cite{Olv,Ste})
but rather that of a Lie module, ${\mathcal G}$ being finite
dimensional as such \cite{CicGae,GaeKlu}. In geometrical terms,
this corresponds to having a finitely generated set of vector
fields in involution {\it \`a la Frobenius}.

As detailed below, our results is of interest mainly when the
invariant manifold whose persistence is considered has a
nontrivial topology. It is well known -- and rather obvious --
that a symmetry of a vector fields maps solutions with a given
topology into solutions with the same topology (we are here
referring to the topology of trajectories for the solutions)
\cite{CicGae,GaeKlu}. The result we give here can also be seen as
a generalization of this to the case where the vector fields as
well as the symmetries depend on control parameters, and moreover
to encompass also the case where the symmetry vector fields also
have (at least a set of) trajectories with compact closure: this
sets further restrictions on the persistence of the resulting
compact invariant manifolds.

\bigskip\noindent
{\bf Acknowledgements.} This work was triggered by the remarks (a
rather long time ago) of prof. J.J. Duistermaat on my previous
work \cite{GaeAOP}. I also gratefully acknowledge useful
discussions with N.N. Nekhoroshev as well as with D. Bambusi, P.
Morando and J. Pejsachowicz.

\section{Statement of the problem}

In this section we describe the general setting of our problem,
i.e. persistence of a $\G$-invariant submanifold $\La_0$ where
$\G$ is an involutory system of vector fields.

We will not use the most general setting, but the most physically
relevant: we deal with a system of parameter-dependent vector
fields on a given manifold $\P$ (the phase space), so that $\N$ is
the direct product of the phase space and a parameter space $\Q$
(a discussion of the general case will be given elsewhere).

Let $\P$ be a smooth manifold of dimension $p$, and $\G = \{ X_1 ,
... , X_d \}$ a set of smooth vector fields in $\P$, which can
depend on external parameters $\eps$, say $\eps \in \R^q$.

We assume $(i)$ that for $\eps = 0$ there is a smooth submanifold
$\La_0 \ss \P$ which is invariant under $\G$, i.e. such that $X_i
: \La_0 \to \T \La_0$ for all $i=1,...,d$.

We assume moreover that there is a tubular neighborhood $V$ of
$\La_0$ such that, for $|\eps|$ small enough, $(ii)$ the vector
fields $\G$ are in involution -- in the Frobenius sense -- in $V$,
and $(iii)$ span a regular distribution in $V$.

Let us briefly recall what these assumptions mean. The $X_i$ are
in Frobenius involution if $[X_i , X_j ] = \s_{ij}^k (x) X_k$,
with $\s_{ij}^k \in \C^\infty (V)$ smooth real functions
(depending also on the external parameters $\eps$) on $V$.

Also, denote by $\D_x \sse \T_x \P$ the distribution associated to
$\G$ at the point $x \in \P$; that is,
$$ \D_x \ := \ \{ v \in \T_x \P \ : \ v = X|_x \, , \, X \in \G \} \ .  $$
(Note that $S \sse V \ss \P$ is $\G$-invariant if, for all $x \in
S$, $\D_x \sse T_x S $.) Then $\D = \bigcup_\P D_x$ is the
distribution on $\T P$ associated to $\G$; this is {\bf regular}
in $U$ if the subspaces $\D_x \sse \T_x \P$ have constant
dimension for $x \in V$.

We would like to identify conditions ensuring that $\La_0$ is part
of a smooth family of $\G$-invariant manifolds $\La_\eps$,
isomorphic to $\La_0$.

It will be convenient to consider the product space $\N = \P
\times \R^q$, and correspondingly $U = V \times \Q$ with $\Q$ a
suitably small neighborhood of the origin in $\R^q$. Our problem
can be studied in $U$. We will also write $\N_c$ for the
intersection of $\N$ with the level manifold $\eps = c$; by
construction, $\N_c \simeq \N_0 = \P$.

\section{The fiber bundle construction}

We will see $U$ as a fiber bundle $(U,\pi,\La_0)$ over $\La_0$,
with contractible fiber. Our problem amounts then to the problem
of identifying conditions which ensure the existence of a family
of $\G$-invariant near-zero sections of this bundle.

We are led to consider a certain (in general, nonlinear)
connection $\nabla$ on this bundle, and covariantly constant
sections of the bundle under $\nabla$.


We define for any point $x \in \La_0$ a local smooth manifold
$\S_x \ss U$ of dimension $s = n-k$ which is transversal to
$\La_0$, with $\S_x \cap \La_0 = \{ x \}$, so that no two such
manifolds intersect, and they define a smooth distribution in $U$
\footnote{If $\N$ is equipped with a riemannian metric, we can
take as $\S_x$ the local geodesic manifold through $x$ orthogonal
to $\La_0$ in $x$.}. We also consider linear local manifolds $S_x$
tangent to $\S_x$ in $x$ (the manifolds $\S_x$ and $S_x$ are
canonically identified by $\nabla$); or to choose $\S_x$ to be
linear, which is fully legitimate \cite{Rue}.

In this way $U$ is a bundle over $\La_0$, and $\S_x$ represents
the fiber through $x \in \La_0$; we denote the corresponding
projection as $\pi$, and write the bundle as $(U,\pi,\La_0)$ with
$\pi^{-1} (x) = \S_x$.

We can also consider $U_0 = U \cap \N_0 = V$, and $\s_x := \S_x
\cap U_0$; these are smooth manifolds of constant dimension $s =
n_p - k$ (and codimension $n_q$ in $\S_x$), and we denote by $s_x
= S_x \cap U_0$ the corresponding local linear manifolds. Then
$U_0$ is also a bundle over $\La_0$, and $\s_x$ represents the
fiber through $x \in \La_0$; we denote the corresponding
projection as $\pi_0$, and write the bundle as $(U_0,\pi_0,\La_0)$
with $\pi_0^{-1} (x) = \s_x$.


The distribution $\D$ associated to $\G$ has constant dimension
$r$ in $U$, by hypothesis. Moreover, $\D$ is tangent to $\La_0$,
and thus transversal to $\S_x$ in $x$ for all $x \in \La$; this
implies that $\D$ is also transversal to $\S_x$ for all points $u
\in \S_x$ sufficiently near to $x$. If $U$ has been chosen to be
sufficiently small -- which we assume from now on -- the
transversality condition is met for all points in $U$. Note that
$\D$ defines a canonical identification between the local
manifolds $\S_x$ and $S_x$ defined above.

We assumed moreover that $k = r$, hence $\D$ defines a (Frobenius
integrable) distribution of horizontal spaces in $U$ and thus a
(in general, nonlinear) {\bf connection} $\nabla$ in $U$
\cite{Chern,Sau}. As this is defined by $\G$,  it will be referred
to as the $\G$-connection in $U$.

The problem of existence of smooth $\G$-invariant manifolds
isomorphic to $\La_0$ and near to it is, with this construction,
translated into the problem of existence of near-zero
$\G$-invariant sections for $U$, i.e. of sections of $U$ which are
invariant under the $\G$-connection $\nabla$.

Note that $\nabla$-invariant sections always exist locally over
any chart $A$ in $\La_0$, due to Frobenius theorem, since $\G$ is
an involutory system; thus the nontrivial part of the problem is
purely global.

As we deal with a small neighborhood of $\La$ we could -- and we
will indeed -- consider the (transverse) linearization of the
$X_i$ around $\La_0$. Correspondingly, we can consider the
(transverse) linearization $\nabla_0$ of the connection $\nabla$.

We stress that $\nabla$ is in general a nonlinear connection,
acting nonlinearly on sections (this is not a problem, as we are
only interested in fixed points under this action), while
$\nabla_0$ is -- by definition -- linear and has an associated
covariant derivative acting linearly on sections. Finally, we note
that, as it follows at once from $\G$ being Frobenius integrable,
the connections $\nabla$ and $\nabla_0$ in $U$ are (locally) flat.

\section{Loops, Poincar\'e-Nekhoroshev map, and monodromy matrices}

Let us fix a reference point $x \in \La_0$, and a loop $\ga$
through $x$ in $\La_0$. Consider then a point $w \in \pi^{-1} (x)
= \S_x$; the loop $\ga$ is lifted by the $\G$-connection to a
curve $\^\ga$ in $U$, which defines a map in $\S_x$.

We associate in this way a (monodromy) map $\M_\ga : \S_x \to
\S_x$ to each loop $\ga$ in $\La_0$. These maps obviously form a
group, the {\it holonomy group} $\Hol_x$ at $x$. When we fix a
basis for the homology of $\La$, we will refer to the maps $\{
\M_{\ga_1} , ... , \M_{\ga_r} \}$ as a set of generators for
$\Hol_x$.

When we consider only contractible loops, the corresponding
subgroup $\Hol^0_x \sse \Hol_x$ is the {\it restricted holonomy
group} at $x$. This is a normal subgroup in $\Hol_x$, and $\Hol_x
/ \Hol^0_x$ is discrete \cite{Nak}. It follows from our assumption
that $\G$ is Frobenius integrable (via the Ambrose-Singer theorem
\cite{AmS,Nak}) that the restricted holonomy group $\Hol^0_x$ is
trivial, $\Hol^0_x = \{ e \}$.

Now, for $x$ an arbitrary reference point on $\La_0$ and $\S_x =
\pi^{-1} x$, $\nabla$-invariant sections of $U$ correspond to
points $w \in \S_x$ which are fixed points for all elements of
$\Hol_x$.

Later on, we will find convenient to focus on the linearization of
$\M_\ga : \S_x \to \S_x$ at $x$; this is a linear operator acting
in the linear space $V = \T_x \S_x \simeq \R^{n-k}$, i.e. a
$(n-k)$-dimensional matrix, called the {\bf monodromy matrix} (at
$x$) for the loop $\ga$. We denote this as $M_\ga$.

In this respect, recall that $\M_\ga$ also defines a map $\P_\ga :
S_x \to S_x$. This will also be called the {\bf
Poincar\'e-Nekhoroshev map} for $\ga$ \cite{GaeAOP,GaeJNMP}. If we
identify $S_x$ with a neighbourhood of the identity in $V$, as can
be done via $\nabla$, then $M_\ga$ is exactly the linearization of
this map.

Monodromy matrices for all loops through $x$ clearly form a group,
called the {\it monodromy group} at $x$, and representing the
linearization of the holonomy group at $x$. Thus we also refer to
this as the linearized holonomy group, and denote it as $\hol_x$.
Given a set of generators for $\Hol_x$, the corresponding
linearized operators will be a set of generators for $\hol_x$.

\Remark{1} As well known, monodromy groups based at different
points are conjugated, and homotopic loops based at the same point
provide the same monodromy maps and matrices.

\section{The difference Poincar\'e-Nekhoroshev map and its linearization}

We have introduced the Poincar\'e-Nekhoroshev map $\P_\ga : S_x
\to S_x$, and its linearization around $\La$, i.e. the monodromy
matrix $M_\ga : V \to V$. By definition, $\P_\ga x = x$; at the
linear level, $x$ corresponds to the origin in $V$, and $M_\ga 0 =
0$ by linearity. Invariant near-zero sections are obtained for the
near-zero $\xi \in V$ such that $M_\ga \xi = \xi$ for all loops
$\ga$.

Instead of considering $\P_\ga$ and $M_\ga$, it is more convenient
(as in the abelian case) to consider the map ${\mathcal R}_\ga :=
I - \P_\ga : S_x \to S_x$ and its linearization (also called the
linearized difference Poincar\'e-Nekhoroshev map) $R_\ga = I -
M_\ga$; note $R_\ga : V \to V$.\footnote{With the construction
considered in \cite{GaeAOP,GaeJNMP} we had to quotient out the
action of $\G$ from the Poincar\'e-Nekhoroshev map; this is not
needed now as we are directly considering the $\G$-connection and
not the flow under specific vector fields $X \in \G$. Actually,
$\M_\ga$ could be seen as a map between local sections (invariant
under the $\G$-connection), see the discussion in
\cite{GaeAOP,GaeJNMP}.}

Let us now focus on a given $x \in \La_0$ and a given loop $\ga
\ss \La_0$ through $x$, and look for $\xi \in \S_x$ such that
${\mathcal R}_\ga \xi = 0 $. By definition, $\xi = x$ satisfies
this, and we are interested in knowing if there is any nearby
$\xi$ satisfying this equation.

This leads to investigate the question of existence of any zero
eigenvalue for the linear map $R_\ga$: if this is the case, the
zero eigenspace corresponds to fixed points of ${\mathcal R}_\ga$
via the implicit function theorem, see below; we stress the
implicit function theorem requires a nondegeneracy condition on
the map.\footnote{We also stress that the argument based on the
implicit function theorem will give sufficient, but not necessary,
conditions.}

If the nondegeneracy conditions are satisfied and there is a
common zero eigenvalue of $R_\ga$ for all $\ga$ through $x$ (zero
eigenvalues of $R_\ga$ correspond to unit eigenvalues of $M_\ga$),
this corresponds to the required near-zero invariant sections, and
hence to invariant manifolds $\La_\eps$ isomorphic and
$\eps$-close to $\La_0$.

Let us make more precise the relation between zeroes of ${\mathcal
R}_\ga$ near the trivial zero $x$ and the zero eigenspace of
$R_\ga$, for a fixed $\ga$. The following lemma follows at once
from the implicit function theorem.\footnote{If we see $R_\ga$ as
a map between local sections, we should use a suitable version
(infinite-dimensional, applying to a space of local sections) of
the implicit function theorem; see e.g. \cite{AmP}.}

\medskip\noindent
{\bf Lemma.} {\it Let ${\mathcal R}_\ga$ and $R_\ga$ be as above;
write  $V = \T_x \S_x \simeq S_x \simeq \R^{n-k}$. Denote the
kernel of $R_\ga$ as $K \sse V$. Assume there is an invariant
complementary space ${\mathcal T}$. Then if $K \simeq \R^m$, there
is a $m$-dimensional manifold $\K \ss \S_x$ of zeroes for
${\mathcal R}_\ga$, and $\T_x \K = K$; for $\eps$ sufficiently
small, all zeroes of ${\mathcal R}_\ga$ within $\eps$ from the
trivial zero $x$ lie on $\K$.}

\Remark{2} Given any loop $\ga$ through $x$, there is a loop
$\ga'$ which is (homotopic to the one) obtained by going $k$ times
round $\ga$; this means that $\hol_x$ will include both $M_\ga$
and all of its powers. If $\mu$ is an eigenvalue of $M_\ga$, there
will be an eigenvalue $\mu^k$ of $M_{\ga '}$; in particular the
presence of an eigenvalue $\mu = e^{i \theta}$ of unit modulus in
the spectrum of $M_\ga$ implies that there will be maps $M_{\ga'}
= (M_\ga)^k \in \hol_x$ with an eigenvalue $\mu' = e^{i k
\theta}$. If $\theta / 2 \pi$ is rational, there is $k$ such that
$\mu' = 1$: thus we will have actually to require that $\mu \not=
\exp (i 2 \pi m/n) $ for all $m,n \in \Z$. It should be stressed
that if $\theta / 2 \pi$ is irrational there will however be
$k,k'$ such that $| k \theta - 2 k' \pi | < \delta$, i.e. $|e^{i k
\theta} - 1| < \delta$, for any $\delta > 0$. This means that in
many contexts -- in particular when discussing bifurcations
\cite{GaeJNMP} -- we should also require that all eigenvalues
satisfy $|\mu_i | \not= 1$ (see e.g. \cite{Arn2} for further
detail). \EOR

The maps $\M_\ga$ leave $U_\eps$ invariant; hence there are
matrices $M_\ga^{(0)}$ and $M_\ga^{(1)}$ such that
$$ M_\ga = \pmatrix{ M_\ga^{(0)} & M_\ga^{(1)} \cr 0 & I \cr} \ ; \ \
R_\ga \ = \ \pmatrix{ I - M_\ga^{(0)} & - M_\ga^{(1)} \cr 0 & 0 \cr} \ . \eqno(1) $$
We denote as $R_\ga^0$ the restriction of the map $R_\ga$ to $V_0$.
By the above formula,
$$ R_\ga^0 \ := \ I - M_\ga^{(0)} \ . \eqno(2) $$

In this case it is immediate to see that the kernel of $R_\ga$ is
provided by
$$ y \ = \  (R_\ga^0)^{-1} \, M_\ga^{(1)} \, z \ , \eqno(3) $$
provided the inverse $(R_\ga^0)^{-1}$ exists (this, of course, is
the same nondegeneracy condition which allowed to use the implicit
function theorem). The condition for this is just that the
spectrum of $M_\ga^{(0)}$ does not include one, i.e. that all the
restricted characteristic multipliers (exponents) satisfy $\mu_i
\not= 1$ ($\b_i \not= 0$). Note that this must hold for all $\ga$
(see also Remark 2 below).

Note also that if $M^{(1)}_\ga = 0$, we are reduced to the
invariant manifold $\La_0$ itself, for all values of $z$.

It should be stressed that in general, formula (3) will give
different $y$ for the same $z$ when we consider different
nonhomotopic loops $\ga$ (a relevant exception is provided by the
case where $\hol_x$ is abelian).

Thus, at difference with the abelian case, we expect that the
condition on spectra of monodromy matrices will not suffice to
ensure the persistence of invariant manifolds; instead, they will
have to be complemented by a "compatibility condition" ensuring
that solutions to (3) for different loops coincide.

On the other hand, if there exists an invariant manifold $\La'$,
intersecting $S_x$ at $\La_x' = \xi$, the point $\xi$ is a fixed
point for $M_\ga$ for any loop $\ga$ through $x$; thus, if $x \in
\ga \cap \ga'$, the commutator $[M_\ga , M_{\ga'}]$ must vanish
when applied to $\xi \in S_x$: the $\xi (y,z)$ identifying
invariant manifolds must belong to ${\rm Ker} ( [h_x,h_x]) \ss V$.

A possible approach, employed in example II below,  to discussing
if solutions to (3) for different loops $\ga$ are compatible is as
follows.

Subdivide $\La_0$ as the union of regions $B_i$, $i = 1,...h$,
each of them with homotopy group $\pi_1 (B_i) = {\bf Z}$; and
denote by $\eta_i$ a homotopically nontrivial loop in $B_i$, so
that the homology of $B_i$ is generated by $\eta_i$. We also write
$M_i$ for $M_{\eta_i}$, $R_i $ for $R_{\eta_i}$, and so on.

Consider then (3) for $\ga = \eta_i$: provided $R_i^0$ is
invertible this identifies, for a given value of $z$, an invariant
manifold $\La_i$ over each of the $B_i$.

A necessary condition for the existence of an invariant manifold
$\La_\eps$ is then that $\La_i = \La_j$ over $B_i \cap B_j$, i.e.
that the invariant manifolds determined in this way by $\eta_i$
and $\eta_j$ do coincide over the intersection of the charts $B_i$
and $B_j$.

If this condition is satisfied over all the nonempty intersection
regions $B_i \cap B_j$, thus determining a possibly invariant
manifold $\^\La (z)$, we should still check this is invariant
under the other generators of the full homology of $\La_0$, i.e.
under the monodromy maps for loops $\ga$ which are not homotopic
to a combination of the $\eta_i$ for $i = 1,...,h$.

\section{The PLN theorem for involutory systems of vector fields}

We have now completed the geometric construction needed for the
Poincar\'e-Lyapounov-Nekhoroshev theorem in the general case, i.e.
for involutory systems of vector fields $\G$ which are not
necessarily a Lie algebra, nor necessarily abelian.

We took care to provide definitions and introduce notations such
to have a statement which looks quite similar to the original one
by Nekhoroshev \cite{Nek,GaeAOP} and a proof (actually,
constituted to a large extent of the discussion conducted so far)
quite similar to the one for the abelian case \cite{GaeAOP}.

\bigskip\noindent
{\bf Theorem 1.} {\it Let $\N$, $\G$ and $\D$ be as in section 2
above. Let $\La_0$ and $U$ be as in sections 2 and 3, with $\La_0$
compact, connected and $k$-dimensional, and $\D$ regular and of
dimension $k$ in $\T U$. Assume moreover that $U$ is foliated into
$\G$-invariant manifolds $U_\eps \simeq \P$ of dimension $n_p$,
with $k < n_p < n$, with $\La_0 \ss U_0$, and $U_\eps$ transversal
to $\pi^{-1} (x)$, so that $V$ admits the decomposition $V =
V^{(0)} \oplus V^{(1)}$, and the monodromy matrices can be written
in the form (1). Let $\eta_1 , ..., \eta_h$ be cycles generating
the homology of $\La_0$, and $\M_j = \M_{\eta_j}$ the associated
monodromy maps; $M_j = M_{\eta_j}$ the associated monodromy
matrices, with $M_j^{(0)}$ their restriction to the space
$V^{(0)}$.
\par\noindent
Then the following are equivalent:

{\tt (i)} There is a $\G$-invariant manifold $\La_\eps \ss
U_\eps$, isomorphic to $\La_0$ and $|\eps|$-near to $\La$, for any
$\eps$ with $|\eps| < \eps^*$;

{\tt (ii)} {\rm (a)} If the loops $\eta_i$ and $\eta_j$ intersect
in some point, the commutator $[M_i,M_j]$ has a nontrivial kernel
in $V_x \simeq \T_x \S_x$; {\rm (b)} The spectrum of any product
of the matrices $M^{(0)}_j$ associated to loops $\eta_j$ such that
$\M_j$ is not the identity does not include points on the unit
circle $S^1 \in \R^2 \simeq {\bf C}$, for all $j = 1,..,h$.}

\bigskip\noindent
{\bf Proof.} It is quite clear that going around loops such that
the associated full monodromy map reduces to the identity will
have no effect on any consideration to follow (which justifies the
specification given in point (ii)-b above), so we can assume all
loops we consider have $\M \not= I$. By the Ambrose-Singer
theorem, this excludes in particular trivial (i.e. contractible)
loops.

Let us first consider loops $\ga$ given by $\eta_j \circ ... \circ
\eta_j = \eta_j^m$ for some loop $\eta_j$. In this case the
monodromy matrix is
$$ M_\ga \ = \ \pmatrix{ [M_j^{(0)}]^m & Z(j,m) \cr 0 & I \cr } $$
with $Z(j,m)$ a matrix whose explicit expression (which could be
given in terms of $M_j^{(0)}$ and $M_j^{(1)}$) is not relevant
here. Thus
$$ R_\ga \ = \ \pmatrix{ I - [M_j^{(0)}]^m & - Z(j,m) \cr 0 & 0 \cr }
\ := \ \pmatrix{ R_j^{(0)} & - Z(j,m) \cr 0 & 0 \cr } \ . $$
The fixed points under $\ga$ will be the kernel of this, i.e. --
with the notation used in section 4 -- will be given by
$$ y \ = \ [R_j^{(0)}]^{-1} \, Z(j,m) \ z \ . $$
In order for $[R_j^{(0)}]^{-1}$ to exist for all $m$, we must
require that eigenvalues of $M_j^{(0)}$ are bounded away from the
unit circle, so that the condition given in the statement is
surely necessary for the existence of $\La_\eps$.

Let us now consider more general loops $\ga$. Any loop $\ga_0$ is
homotopic (suitably choosing the base point, see also Remark 1) to
a loop $\ga$ which can be written as
$$ \ga \ = \ \eta_{j_1} \circ \ ... \ \circ \eta_{j_\nu} $$
for some sequence of $\nu \in (1,...,n) $.
The associated monodromy matrix is
$$ M_{\ga} \ = \ \pmatrix{ M_{j_1}^{(0)} ... M_{j_\nu}^{(0)} &
\beta \cr 0 & I \cr } \ = \ \pmatrix{\a & \b \cr 0 & I \cr} $$
with $\beta$ a complicate expression we do not need to write
explicitly.

As usual, we look at the matrix $R_{\ga} = I - M_{\ga}$, and we
are concerned with the invertibility of $ R_\ga^{(0)}$, which is
now given by $R_\ga^{(0)} := I - \a$. Thus, in order to have an
invariant manifold we have to require that the spectrum of $\a$
does not contain the unity. This is precisely {\tt (ii)-b}.

Finally, note that we need invariant manifolds associated to
different loops going through the same point do coincide; this is
precisely {\tt (ii)-a}, and the proof is complete. \EOP

\bigskip\noindent
{\bf Corollary 1.} {\it Under the assumptions and with the
notation of theorem 1 above, assume moreover that the
$\G$-invariant manifold $\La_0$ has trivial homotopy group, $\pi_1
(\La_0) = I$. Then there is $\eps^* > 0$ such that in the tubular
neighborhood of $\La_0$ of radius $\eps^*$ there is a
$\G$-invariant manifold $\La$, foliated into $\G$-invariant
manifolds $\La_\eps$ isomorphic to $\La_0$.}

\bigskip\noindent
{\bf Proof.} If $\pi_1 (\La ) = \{ e \}$, all loops $\ga$ in $\La$
are contractible; hence $\Hol_x$ reduce to $\Hol^0_x$. From the
Frobenius integrability of the distribution $\G$ it follows that
$\nabla$ is flat, and the Ambrose-Singer theorem \cite{AmS,Nak}
guarantees that $\Hol = I$. \EOP

Let us present some short remarks on the results obtained above.

{\bf 1.} It might be appropriate to stress that in theorem 1 the
condition {\tt (ii)-b} on the spectrum of monodromy matrices is
(once {\tt (ii)-a} is granted) sufficient but not necessary to
guarantee the existence of a continuous family of $\G$-invariant
manifolds, i.e. {\tt (i)}. To see this, just think of the case
where the monodromy operators of all loops $\ga$ are just $\M_\ga
= I$.

{\bf 2.} We note that when $\La_0$ is of codimension one in $U_0$
(e.g., if it is of codimension one in phase space), $M^{(0)}_\ga$
are just numbers, and it is very easy to check the conditions
given in theorem 1. Similarly, if it happens that the
$M_{\eta_j}^{(0)}$ commute (albeit the full $M_{\eta_j}$ may not
commute) it is easy to check that condition.

{\bf 3.} Note that if $\La_0$ is the topological product $\La_0 =
\^\La_0 \times B$ of a topologically nontrivial manifold $\^\La_0$
and a contractible manifold $B$, then by lemma 3 it suffices to
consider $\^\La_0$ (as it also follows from the role of homology
groups in our discussion). If $\La_0$ is contractible, then our
result is trivial.

{\bf 4.} It should be stressed that our discussion encompasses
cases where -- by topological reasons -- any vector field on
$\La_0$ necessarily has fixed points, so that the considered
invariant manifold is necessarily not minimal. By a naive parallel
with the torus case one could think that in this case the
invariant manifold breaks down under perturbations, but actually
the same topological constraints guarantees its persistence. In a
way, only degenerations which are not enforced by topology are
dangerous for persistence.

{\bf 5.} Finally, we note that the abstract results obtained here
are qualitatively equivalent to those holding for tori and
commuting vector fields \cite{Nek,BamGae,GaeAOP,GaeJNMP}; but now
checking that the hypotheses hold in concrete situations will in
general be quite more difficult than in the abelian case.

\section{The coordinate picture}

We have so far conducted our discussion in rather abstract terms,
in order to make clear the geometric content of our result.
However, in order to use it in concrete situations it is
convenient to have a formulation in local coordinates as well;
this is the aim of the present section.


We will now introduce local coordinates in $U$ and in $U_0$.
Consider a local chart $B_i \ss \La_0$ with local coordinates
$\phi$; we naturally associate to this a chart $A_i = \pi^{-1}
(B_i)$ in $U$. The natural local coordinates for this will be
$(\phi,w)$ where $\phi$ are local coordinates in $B_i$, and
$w=(w^1,...,w^S)$ are coordinates on the fiber $\pi^{-1} [x
(\phi)]$.

It is convenient to choose these as $w = (y,z)$ where $y =
(y^1,...,y^{p-k})$ are coordinates in $\s_x$ and $z = (z^1 , ... ,
z^a )$ are coordinates in the parameter space $\Q$. (For ease of
later notation, we choose coordinates $(y,z)$ on $\S_x$ such that
$|y^i|,|z^a| \le \de$.)

The vector fields $X_a$ generating $\G$ will be written in these
coordinates as
$$ X_a \ = \ \b_a^i {\pa \over \pa \phi^i} \ + \ f_a^k {\pa \over \pa w^k } \ . $$

The linearized connection will be generated by the linearizations
of the $X_a$ around $\La$; this amounts to replacing $\b^i_a$ and
$f^k_a$ by their linearization\footnote{We recall that
linearization is always to be meant in transversal (to $\La_0$)
sense; this means linearization in the $w=(y,z)$ coordinates, but
not in the $\phi$ ones.} at $\La$:
$$ \begin{array}{rl}
X_a^0 \ =& \ \[ \b_a^i (\phi , 0 ) + \({\pa \b_a^i \over \pa w^s} (\phi , 0 ) \) w^s \] \, {\pa \over \pa \phi^i} \ + \\
 & \ + \ \[ f_a^k (\phi, 0 ) + \( {\pa f_a^k \over \pa w^s} (\phi , 0 ) \)  w^s \] \, {\pa \over \pa w^k } \ . \end{array} $$

As stressed above, we can just consider the action of the {\it
linearized holonomy group} at $x$, made of $M_\ga$ for all loops
$\ga$ in $\La_0$. Its explicit construction is quite standard, but
we discuss it briefly below in order to fix notation.


The nonlinear connection $\nabla$ provides a lift of the ordinary
derivative $\pa_\phi$ along $\ga$, which is in general a nonlinear
operator (and thus not a proper covariant derivative). This is
written in the $(\phi , w)$  coordinates as
$$ \nabla_\phi \ = \ \pa_\phi + A^i (\phi,w) {\pa \over \pa w^i} \ , $$
where $A^i : \Ga \to \R$ is in general a nonlinear function of
$w$;  as $\La_0$ is $\G$-invariant, it must be $ A^i (\phi,0) =
0$. Linearizing around $\ga$ we have a covariant derivative
$$ D_\phi \ = \ {\pa \over \pa \phi} \ + \ (L^i_j w^j) {\pa \over \pa w^i} $$
where the matrix $L$, given by $ L^i_j := (\pa A^i / \pa w^j)
(\phi , 0)$, is a function of $\phi$ alone. The evolution of
$\^\ga (t) = ( \phi (t) , w (t) )$ along the loop coordinate $t$,
i.e. along $\phi = \phi_0 + 2 \pi t$ is described by $ {\dot \phi}
= 2 \pi$, ${\dot w} = L w$; needless to say, the solution to these
is
$$ \phi (t) = \phi_0 + 2 \pi t \ \ ; \ \ w (t) = \exp \[ \int_0^t L[\phi(\tau)] \d \tau \] \ w_0 \ . $$

The function $\phi (t)$ is (as obvious by construction) periodic
of period one; as $L = L (\phi)$, the matrix $L$ is also
$t$-periodic of period one, and we are thus considering a
time-periodic vector equation for the vector variable $w (t) \in
\R^{n-k}$. We can then invoke Floquet theorem (see e.g.
\cite{Gle,Rue,Ver}).

\medskip\noindent
{\bf Proposition.} (Floquet theorem) {\it The fundamental solution
matrix $\Psi (t)$ for ${\dot \xi} = L (t) \xi$ with $L$ a matrix
$T$-periodic in $t$ can be written as $\Psi (t) = P(t) \exp[ B t]$
with $P(t)$ a $T$-periodic matrix, and $B$ a constant matrix. In
particular, $\Psi (t+T) = \Psi (t) M$, with $M = \exp [ B T]$.}
\medskip

The matrix $M$ in the statement above is precisely the monodromy
matrix for the given loop $\ga \ss \La$. The eigenvalues $\mu_1 ,
... , \mu_{n-k}$ of $M$ are the {\it characteristic} (or Floquet)
{\it multipliers}; if $\nu_i$ are the eigenvalues of $B$ we have
$\mu_i = e^{\nu_i T}$, and the $\nu_i = \log (\mu_i T)$ are the
{\it characteristic} (or Floquet) {\it exponents}.

The information provided by the spectrum of $M$ is best used by
passing to coordinates $\eta$ via $\xi = P (t) \eta$; in these the
evolution equations read ${\dot \eta} = B \eta$, so that $\eta (t)
= e^{Bt} \eta_0$ and of course $ \eta (T) = M  \eta_0$.

It is worth stressing that the unit vector field $Y$ tangent to $\^\ga
(t)$ in $t$ can always be written in the form $Y = \a_i (t) X_i$,
but in general it is not possible to find a $\ga' \simeq \ga$ such
that one can choose the $\a_i (t)$ as constant, contrary to the
abelian case. (However the loop $\ga$ can be deformed into a
piecewise smooth path $\ga '$ homotopic to $\ga$ on which the
$\a_i (t)$ are piecewise constant.)
Thus in the case where $\G$ is
a Lie algebra, $Y$ does not belong to the algebra $\G$, but to the
module over $\C^\infty (\ga')$ generated by $\G$. In the
terminology of field theory, we should consider the {\bf gauge algebra}
modelled on $\G$.


With the notation introduced earlier on in this section, we will
write
$$ \nabla_\phi \ = \ \pa_\phi + f^i (\phi,y,z) {\pa \over \pa y^i} + h^a (\phi,y,z) {\pa \over \pa z^a} \ . $$
The $\G$-invariance of $\La_0$ entails that $ f^i (\phi,0,0) = h^a
(\phi,0,0) = 0 $. Linearizing around $\ga$ we get
$$ D_\phi \ = \ {\pa \over \pa \phi} \ + \ (F^i_j y^j + G^i_a z^a) {\pa \over \pa y^i} \ + \ (H^a_j y^j + K^a_b z^b ) {\pa \over \pa z^a}   $$
where the matrices $F,G,H,K$ are functions of $\phi$ alone, and of
course we have defined these by $ F^i_j (\phi ) := (\pa f^i / \pa
y^j)_{\La_0}$, $G^i_a (\phi ) := (\pa f^i / \pa z^a)_{\La_0}$,
$H^a_j (\phi ) := (\pa h^a / \pa y^j)_{\La_0}$, $K^a_b (\phi ) :=
(\pa h^a / \pa z^b)_{\La_0}$.

In other words, the evolution of the coordinates $(\phi , y,z)$
with the curvilinear coordinate $t$ along the loop $\ga$ is given
by
$$ \cases{
{\dot \phi} = 1 & \cr
{\dot y^i} = F^i_j (\phi ) y^j + G^i_a (\phi ) z^a & \cr
{\dot z^a} = H^a_j (\phi ) y^j + K^a_b (\phi ) z^b \ . &  \cr} $$

The $\G$-connection leaves all the level sets $z=c$ also
invariant. This implies in turn that $h^a (\phi;y,z) = 0$, hence
$\M_\ga : \s_x \to \s_x$. At the linear level, we have $H^a_j
(\phi) \equiv 0 \equiv K^a_b (\phi)$ for all loops $\ga$, and
$M_\ga : V_0 \to V_0$.

In other words, for any $\ga$ we can write the monodromy matrix in the form
$$ M_\ga \ = \ \pmatrix{ M_\ga^{(0)} & M_\ga^{(1)} \cr 0 & I \cr} \ . $$
Hence, $q$ of the Floquet multipliers will be trivially given by
$\mu_i = 1$, $i = 1,...,q$. The corresponding Floquet exponents
are $\nu_i = 0$. We can thus, under the local foliation
hypothesis, restrict the matrix $M_\ga$ to the $(p-k)$-dimensional
space $V_0$, obtaining the matrix $M_\ga^{(0)}$ considered above;
we refer to this as the {\bf restricted monodromy matrix} for the
loop $\ga$. Its eigenvalues $\a_i$ will be called the {\it
restricted characteristic} (Floquet) {\it multipliers}, and $\b_i
= \log (\a_i)$ will be the {\it restricted characteristic}
(Floquet) {\it exponents}. The restricted multipliers (exponents)
do of course encode all the nontrivial information.

\section{Example I}

Let us consider $\P = \R^3$ with spherical coordinates
$(\phi,\theta,r)$, $\Q = \R$ with coordinate $\la$, and vector
fields
$$ \begin{array}{rl}
X \ :=& \ f_1 (r,\la) \, \pa_\phi \, + \, f_2 (r,\la) \, \pa_\theta \, + \, f_3 (r,\la ) \, \pa_r \ , \\
Y \ :=& \ [\a_1 (\la) + \a_3 (\la) f_1 (r,\la)] \, \pa_\phi \, + \\
 & \ + \, [ a_3 (\la) + a_3 (\la ) f_2 (r,\la)] \, \pa_\theta \, + \, [a_3 (\la ) f_3 (r,\la )] \, \pa_r \ , \end{array} $$
where all functions are smooth in their arguments. These commute
for all values of $r$ and $\la$, $[X,Y] \equiv 0$\footnote{Hence,
if desired, we can see one of them as defining a dynamics in
$\R^3$, and the other as a symmetry of this dynamics.}. Their
cross product is given by
$$ X \times Y \ = \  - \a_2 (\la) f_3 (r,\la) \pa_\phi \, +
\a_1 (\la) f_3 (r,\la ) \pa_\theta \, + \a_2 (\la ) f_1 (r,\la) -
a_1 (\la ) f_2 (r,\la) \pa_r \ ;  $$ thus, the distribution $\G$
spanned by $X$ and $Y$ is regular and two dimensional in $Q_0 \sse
\Q$ provided the three components of this do not vanish
simultaneously in any point of $Q_0$, i.e. $|| (X \times Y ) ||
\not= 0$.

E.g. for $f_3 (r,\la) = r - k(\la)$ this is the case provided
$$ \a_2 (\la ) f_1 [k(\la),\la] - a_1 (\la ) f_2 [k(\la),\la] \not= 0 \ \ \forall \la \in Q_0 \ . $$
In this case there is obviously an invariant sphere $S^2$ of
radius $k(\la )$ for all values of $\la$ such that $k(\la ) > 0$.

Suppose now that $|| (X \times Y ) || \not= 0$ is satisfied on the
sphere of radius $r_0$, which we denote as $\La_0$ (and therefore
in a tubular neighbourhood $U$ of it in $\N = \P \times \Q$ of
sufficiently small radius $\eps$), and that for $\la = 0$ the
function $f_3$ satisfies $f_3 (r_0,0) = 0$, $(\pa_r f_3) (r_0,0)
\not= 0$, so that $\La_0$ is an isolated invariant manifold for
the distribution $\G$.

Our discussion, and in particular Corollary 2, guarantee that for
$|\la| < \eps$ there is a $\G$-invariant manifold isomorphic to
$\La_0$ (actually, for our simplyfying choice of the vector fields
this will also be a $S^2$ sphere), as the homotopy group $\pi_1
(S^2)$ is trivial.

It has to be noted that this can also be seen without making use
of our result, as a simple consequence of the implicit function
theorem applied to $f_3 (r,\la)$; which is not surprising as,
after all, our results were also based on that theorem.

\section{Example II}

Let us consider $\P = R^3$, $\Q = \R$, and let $\La_0$ be the
two-dimensional ``double torus'' (or ``two-holes pretzel''), see
fig.1; as well known, there is no way to have a nowhere zero
vector field on it, so we will have more vector fields, still
providing a two-dimensional distribution.

\begin{figure}
\includegraphics[width=300pt]{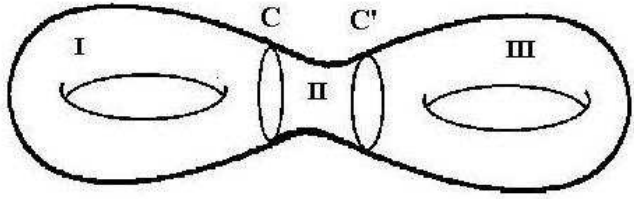}
\caption{The double torus. The circles $C$ and $C'$ represent the
borders between regions I, II and III (see also fig.2).}
\end{figure}

Our purpose here is to build an example showing in explicit terms
the validity of our general result in this special case. It will
also be clear how to extend this example to the case of a pretzel
with $g$ holes.


We decompose $\La_0$ into three (flat) regions $B_i$ as suggested
in fig.1; two of these are ``tori with a cut'', while the central
one is a cylinder. This is also illustrated in fig.2, where it is
also shown how the three regions are glued together. (One could as
well decompose it into two regions isomorphic to $B_1$ and $B_3$;
we use the three-decomposition as this makes more clear the
monodromy computation below.) With a slight abuse of notation, we
will refer to the $B_i$ equipped with a coordinate system as
``charts'', although each of these is actually the union of
charts, and to their union as an atlas.

\begin{figure}
\includegraphics[width=300pt]{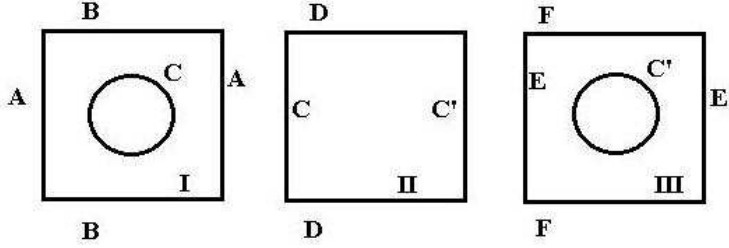}
\caption{The decomposition of the double torus into three flat
regions. The letters show which lines should be identified.}
\end{figure}

We take coordinates $(x_i,y_i)$ on each of the $B_i$. We can take
the origin of the coordinate systems at the center of the
rectangle representing each region; we take $x_i,y_i$ to range
from $-\pi$ to $\pi$ for $i=1,3$, and state for ease of discussion
-- but with no loss of generality -- that for the central chart
$x_2 \in [-1,1]$, $y_2 \in [-\pi , \pi]$.

The $B_i$ are immediately extended to charts (in the same sense,
i.e. with the same abuse of notation, as above) and to an atlas on
a tubular neighbourhood $U_0 \ss \P \times \Q$ of $\La_0$ of width
$\eps$; the coordinates on the chart $A_i$ built over $B_i$ will
simply be $(x_i,y_i,z_i;\lambda)$.


We define two vector fields on each of the $A_i$, given in local
coordinates\footnote{Note that more precisely we should add to
each vector field a smoothing factor, being 1 out of the
transition regions, and smoothly sending the vector fields to zero
at the border of their domain of definition. However, introduction
of these would merely add to commutators some factors proportional
to vector fields, so the new terms would not change the module
structure: for ease of notation (and computation) we will just
drop these.} by
$$ X_i \ := \ {\pa \over \pa x_i } \ + \
\phi_i (x_i,y_i,z_i;\la) {\pa \over \pa z_i}  \ ; \
Y_i \ := \ {\pa \over \pa y_i } \ + \
\psi_i (x_i,y_i,z_i;\la) {\pa \over \pa z_i} \ . \eqno(4) $$
Note that in order for $\La_0$ to be invariant under these, we
must require that $\phi_i$ and $\psi_i$ vanish on $\La_0$, i.e.
that $ \phi_i (x_i,y_i,0;0) = 0$, $\psi_i (x_i,y_i,0;0) = 0 $.

If we linearize in the ``vertical coordinates'' $z_i$ and in
$\la$,  with $f_i (x_i , y_i ) := \( \pa \phi_i / \pa z_i
\)_{\La_0}$, $g_i (x_i , y_i ) := \( \pa \psi_i / \pa z_i
\)_{\La_0}$, and similarly $\^f_i (x_i , y_i ) := \( \pa \phi_i /
\pa \la  \)_{\La_0}$, $\^g_i (x_i , y_i ) := \( \pa \psi_i / \pa
\la  \)_{\La_0}$, we get $$ X_i \ := \ {\pa \over \pa x_i } \ + \
\( f_i z_i  + \^f_i \lambda  \) {\pa \over \pa z_i}  \ ; \ Y_i \
:= \ {\pa \over \pa y_i } \ + \ \( g_i z_i + \^g_i \lambda  \)
{\pa \over \pa z_i} \ . \eqno(4') $$

A particularly simple but nontrivial choice (in which the vector
field are chosen to be linear in $z$ and $\la$), corresponding to
$f_1 = \^f_1 = f_3 = \^f_3 = 0$, $f_2 = q(x)$, $\^f_2 = - q(x)
b_2$, $g_1 = g_2 = g_3 = 1$, $\^g_i = - b_i$, is the following:
$$ X_i := {\pa \over \pa x_i } + \de_{i,2} \[ q(x) (z_i - b_i \la)  \] {\pa \over \pa z_i} \ , \
Y_i := {\pa \over \pa y_i } + \[ z_i - b_i \la \] {\pa \over \pa z_i} \ . \eqno(5) $$
Here the $b_i$ are real constants, while $q(x)$ is a smooth function.

Let us now consider the transition regions $A_i \cap A_j$. For the
sake of concreteness -- and in order to introduce a notational
simplification -- let us just focus on $A_1 \cap A_2$. We will
write $ x_1 = \xi$, $y_1 = \eta$, $z_1 = \zeta$; $x_2 = x$, $y_2 =
y$, $z_2 = z$.

It is quite clear that we can just take $\z = z$, and that $\xi =
\xi (x,y,z) $, $\eta = \eta (x,y,z)$ can be taken to be
independent of $z$; so $z = \zeta$ (and of course the parameter
$\lambda$) will just drop from our discussion of transition
functions: these can be discussed in $B_i \cap B_j$.

It is convenient to use polar coordinates $(\rho,\vth )$ in $B_1$.
For a point $(\xi,\eta) \simeq (x , y) \in B_1 \cap B_2$, with our
choice for the $x$ range and origin, $\rho = (\xi^2 +
\eta^2)^{1/2}$ is just $ \rho = r_0 + (1 + x)$, where $r_0$ is the
radius of the excluded circle in $B_1$ (recall $0 < r_0 < 1$); as
this function will appear often in the following, we will denote
it by $ h(x) \ := \ 1 + r_0 + x$.

As for the angle, we can just take $\vartheta = y$. Combining
these with $\xi = \rho \cos (\vth ) $ and $\eta = \rho \sin (\vth
)$, we obtain at once the direct and inverse transition functions:
$$ \begin{array}{ll}
\xi \ = \ h(x) \,  \cos (y ) \ ;& \ \eta \ = \ h(x) \,  \sin (y ) \\
x \ = \ \sqrt{\xi^2 + \eta^2} \, - \, (1 + r_0 ) \ ;& \ y \ = \ \arctan(\eta / \xi ) \ . \end{array} $$

With the choice (5) for the vector fields, and the above, we get
$$ \begin{array}{ll}
X_1 := \[\cos (y )\] {\pa \over \pa x}  - \[{\sin (y ) \over h(x)}\] {\pa \over \pa y} ,&
Y_1 := \[ \sin (y ) \] {\pa \over \pa x } + \[{\cos (y ) \over  h(x)}\]  {\pa \over \pa y} + \[ z - b_1 \la \] {\pa \over \pa z} \ ; \\
X_2 := {\pa \over \pa x } + \[ q(x) (z - b_2 \la ) \] {\pa \over \pa z} ,&
Y_2 := {\pa \over \pa y } + \[ z - b_2 \la \] {\pa \over \pa z} \ . \end{array}  $$


Let us now consider the commutation relations. It is easy to see
that the condition (no sum over repeated indices from now on)
$$ [X_i , Y_i ] \ = \ \s_1^{(i)} X_i + \s_2^{(i)} Y_i $$
which must be satisfied in each of the $A_i \backslash [ \cup_{j
\not=i} (A_i \cap A_j ) ]$ for the vector fields to be in
involution, actually imposes $\s_1^{(i)} = \s_2^{(i)} = 0$, i.e.
$[X_i , Y_i] = 0$ (so that again we can see one of $X$ and $Y$ as
defining a dynamics, and the other as a symmetry of this
dynamics). It is also easy to check that these are satisfied with
our choice (5) for the vector fields.\footnote{More in general,
$[X_i,Y_i] = 0$ requires, with reference to (4'), that $(\pa f_i /
\pa y_i) = (\pa g_i / \pa x_i)$ and $\^f_i g_i  - \^g_i f_i = (\pa
\^f_i / \pa y_i )  - (\pa \^g_i / \pa x_i )$. Defining on each
chart the one-forms $ \a_i := f_i \d x_i + g_i \d y_i $
(associated to derivatives in $z$) and $\^\a_i  := \^f_i \d x_i +
\^g_i \d y_i$ (associated to derivatives in $\la$), these are also
rewritten as $ \d \a_i = 0$ and $\d \^\a_i = \a_i \w \^\a_i$. Note
that as the $B_i$ are not contractible, we are {\it not } required
to have $\a_i = \d \Phi_i$; actually the most interesting case
will be the one where $\a_i \in H^1 (B_i)$.}

Let us now discuss the commutation relations in the transition regions: in each of these four fields are present, and they should be in involution. Consider, for definiteness, $A_1 \cap A_2$. By the Frobenius condition, we must require e.g. that there are functions $\s_i (x,y)$, $\mu_i (x,y)$, smooth in $A_1 \cap A_2$, such that
$$ [X_1 , X_2 ] \ = \ {\mathcal S} \ := \ \s_1 X_1 + \s_2 X_2 + \mu_1 Y_1 + \mu_2 Y_2 \ ; \eqno(6) $$
similar equations also hold for the other commutators.

One can check by explicit computations (see the appendix) that
with our choice (5) for the vector fields, eq.(6) and those for
the other relevant commutators admit a well defined solution under
the condition that $ \chi (x,y) := h(x) [q(x) - \sin (y) ]$ does
not vanish in $B_1 \cap B_2$, i.e. for $r_0 < x + 1 < r_1$,
equivalently  $-1 < x < -1 + \delta$.

As $h(x) := (1 + x + r_0) > 0 $ in $B_1 \cap B_2$, we have to require that in this region $ q(x) \not= \cos (y)$.
We can e.g. require that for $-1 < x < -1 + \delta$ the function $q(x) $ satisfies $| q(x)| > 1$. This leaves ample freedom of choice for that function.

Finally, we note that if $f_i , g_i$ in (4') are not zero, then there is no manifold near to $\La_0$ (and homeomorphic to $\La_0$) which is also invariant under these vector fields (this condition is relevant, in particular, if we are interested in bifurcations from invariant manifolds; see also the discussion in \cite{GaeJNMP} for bifurcation from Poincar\'e-Lyapounov-Nekhoroshev invariant tori).


In order to illustrate our result, we have to consider the
monodromy matrices associated to cycles providing a base for the
homology of $\La_0$. We consider the cycles in $\La_0$ illustrated
in figs.3 and 4.

\begin{figure}
\includegraphics[width=300pt]{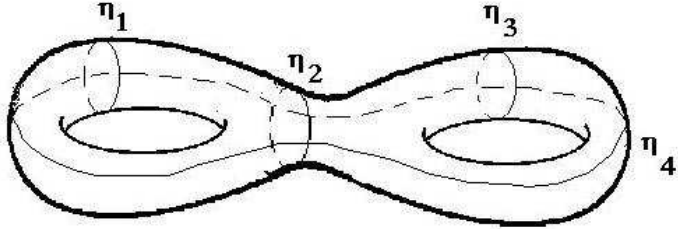}
\caption{The four cycles, providing a basis for the homology of
the double torus.}
\end{figure}

\begin{figure}
\includegraphics[width=300pt]{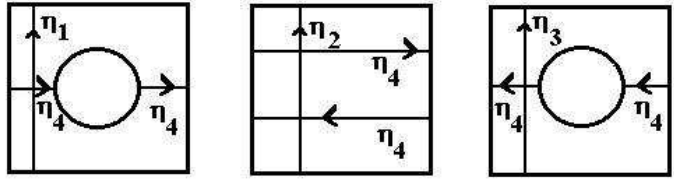}
\caption{The four cycles, providing a basis for the homology of
the double torus, here shown using the same representation as in
fig.2.}
\end{figure}

Computation of the monodromy matrices for the cycles $\eta_1 ,
\eta_2 , \eta_3$ is immediate: each of these lies on a single
region $B_i$, and moreover involve only $Y_i$. Along these paths,
parametrized with $t \in [0,2 \pi]$, we have (dropping the
subscript $i$) $ d x / d t = 0$, $d y / d t = 1$, $d z / d t = z -
b \la$, $d \la / d t = 0$; the solution for $z (t)$ (with $z(0) =
z_0$) is given by
$$ z(t) \ = \ e^{t - t_0} \, z_0 \ + \ b \, \(1 -  e^{t-t_0} \) \, \la \ . $$
Therefore, writing $\kappa = \exp (2 \pi)$, the monodromy matrix $M_i$ (coinciding with the monodromy map $\M_i$, as this is linear) and the associated $R_i$ are given by
$$ M_i \ = \ \pmatrix{ \kappa & (1 - \kappa ) b_i \cr
0 & 1 \cr} \ ; \ R_i  \ = \ \pmatrix{1 - \kappa & (\kappa -1) b_i \cr 0 & 0 \cr} \ . $$
The invariant local manifolds (under transport by $\nabla \equiv \nabla_0$ along $\eta_i$) are provided, as obvious from the explicit form of $Y_i$, by
$ z = b_i \la$.
Needless to say, these local manifolds glue together into a global manifold if and only if $b_1 = b_2 = b_3 = b$.

Let us now consider the path $\eta_4$ and the associated monodromy
matrix $M_4$. The path $\eta_4$ is decomposed as $\tau_4 \circ ...
\circ \tau_1$, see figure 4. With our choice of $X_i$, only the
parts $\tau_1$ and $\tau_3$ will contribute to the monodromy map;
that is, ${\dot z}  =0$ for $\tau_2$ and $\tau_4$. As for $\tau_1$
and $\tau_3$, here ${\dot x} = 1$, ${\dot z} = q(x) [z - b \la]$.
We have therefore to solve $ \d z / \d x = q(x) [ z  -  b \la ]$;
writing in full generality $q(x) = Q' (x)$, this has solution
$$ z (x) \ = \ b \la \ + \ \exp \[ Q(x) - Q(x_0 ) \] \, (z_0 - b \la ) $$ for initial datum $z (x_0 ) = z_0$. Note this, like the equation for $z(x)$ itself, is independent of $y$.

It is easy to see from the above expression and simple algebra (or
directly from the $y$-independence) that the monodromy map $M_4 :
(z_0,\la) \mapsto (z_2,\la)$ is the identity. (It should be
stressed that this is true not only of the linearized map, but of
the full monodromy map, whenever $\phi (x,y,z;\la) = \phi
(x,y+\pi,z;\la)$ or  does not depend on $y$. See, in this respect,
the third remark at the end of section 6.)

In conclusion, we have checked that -- as actually obvious from
the form of the vector fields -- the smooth family of manifolds
identified by $z = b \la$ is invariant under $\G$.

\section{Example III}

We will now provide a framework where the situation considered in
example II is met in practice.

Hamiltonian systems with nontrivial topology of the relevant
energy manifold have been studied by a number of authors, see e.g.
\cite{Koz}, and \cite{Gav} for a recent contribution focusing on
isochronous hamiltonian systems; isochronous systems on Riemann
surfaces extremely robust under perturbations have been considered
by Calogero \cite{Cal}. Here we discuss simpler systems; the
quantum version of these is studied in \cite{Cheng}.

We consider a system (not necessarily hamiltonian) describing a
point particle in $R^4$, with cartesian coordinates $(x,y,z,w)$.
We also introduce "bi-polar" coordinates $(r_1,\vth_1,r_2,\vth_2)$
by
$$ \begin{array}{ll}
x = r_1 \cos \vth_1 \ , \ & y = r_1 \sin \vth_1 \ , \\
z = r_2 \cos \vth_2 \ , \ & w = r_2 \sin \vth_2 \ . \end{array} $$
With $r_j = \sqrt{I_j}$ and $x= p_1$, $y=q_1$, $z=p_2$, $w=q_2$,
we would be in a hamiltonian framework and $(I,\vth)$ be
action-angle coordinates.

In these terms, the phase space is described as
$$ {\mathcal P} \ = \ \toro^2 \times \R_+^2 \ . $$

Let us now introduce in $\R^4$ a solid cone ${\mathcal C}$ of
angle $\Phi$, with vertex in the origin and surface $\pa {\mathcal
C}$ described in the bi-polar coordinates by
$$ \vth_1^2 \, + \, \vth_2^2 \ = \ \Phi^2 \ . \eqno(7) $$
This will be a solid surface on which the point particle bounces
elastically. Thus, the accessible phase space ${\mathcal P}_a =
{\mathcal P} \backslash {\mathcal C}$ will be described in
bi-polar coordinates as
$$ {\mathcal P}_a \ = \ \toro^2_\Phi \times \R_+^2 \ , $$
where $\toro^2_\Phi$ is the torus $(\vth_1,\vth_2) \in S^1 \times
S^1$ without the two-dimensional disk $D^2_\Phi$ of radius $\Phi$,
see (7), $\toro^2_\Phi = \toro^2 \backslash D^2_\Phi$.

Note that each time the particle hits on the surface $\pa
{\mathcal C}$, the component $p_0$ of its momentum in the
direction orthogonal to the surface is reflected into $- p_0$,
while other components of momentum -- and a fortiori its position
-- are continuous. The motion takes then place normally until next
hit on the surface, when again $p_0$ is reflected into $- p_0$ and
so on. We can thus represent the motion as taking place on a
double covering ${\mathcal M}$ of ${\mathcal P}_a$, the two sheets
of this Riemann surface\footnote{The complex structure of this can
be chosen in a number of ways, e.g. by introducing complex
coordinates $\zeta_1 = x + i y$, $\zeta_2 = z + i w$, or even
seeing this as the product of $\R_+^2$ by a complex surface with
complex coordinate $\eta = \vth_1 + i \vth_2$.} merging precisely
on $\pa {\mathcal C}$. Needless to say, this construction is just
a manifestation of the classical Schwarz reflection principle.

We would now like to present some remarks concerning this
construction.
\begin{itemize}
\item {\bf (a)} First of all, note that the dynamics is singular
-- and actually not uniquely defined -- on $S^1_0 := \pa {\mathcal
C}$; indeed, on these points the dynamics is defined by continuity
and thus in particular it depends on the direction of approach
(from one or the other of the two sheets of the double covering
${\mathcal M}$).

\item {\bf (b)} In view of our construction, it is entirely
natural to consider a two-charts decomposition of $\La_0$, each
chart being isomorphic to ${\mathcal P}_a$ and corresponding to
one sheet of the double covering ${\mathcal M}$; see fig.5. This
also prompts for a different choice of the basis cycles for the
homology of $\La_0$, see fig.6.

\item {\bf (c)} Our construction implies the vector fields to be
considered do naturally satisfy an antisymmetry condition with
respect to $S^1_0$. This simplifies the computation of monodromy
matrices, in particular if we adopt an appropriate choice for the
basis cycles, each of them lying in a single chart of the double
covering: indeed, it suffices then to compute the monodromy for
cycles belonging to a given chart.
\end{itemize}

\begin{figure}
  \includegraphics[width=200pt]{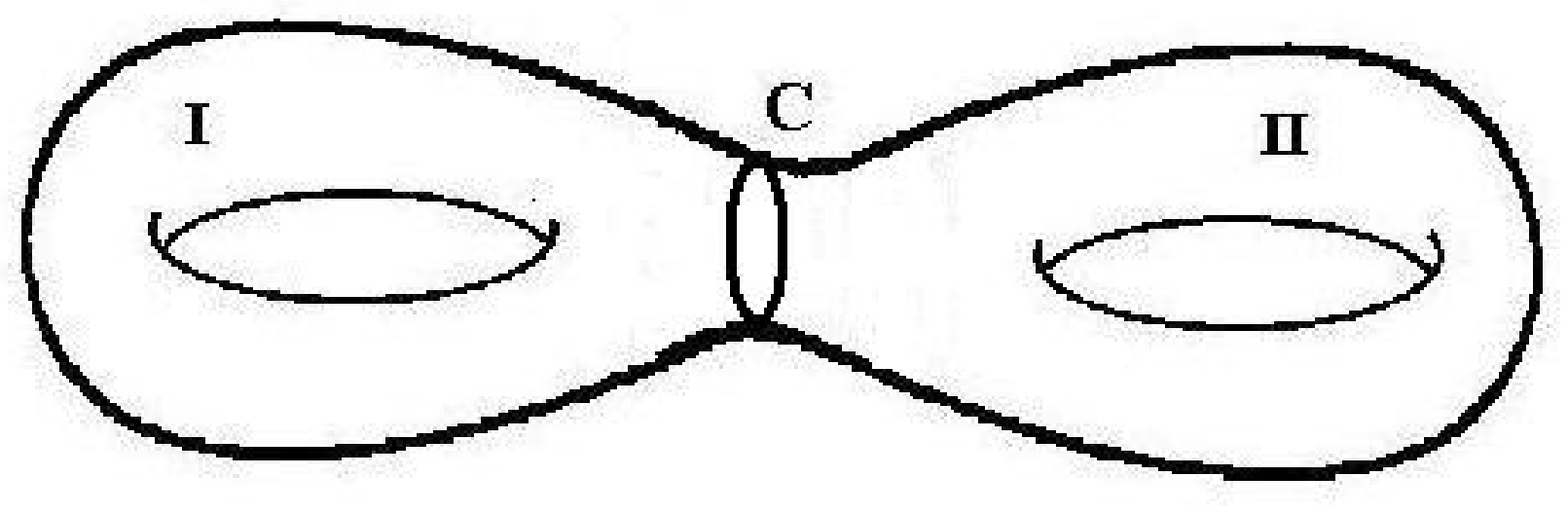}\ \
  \includegraphics[width=150pt]{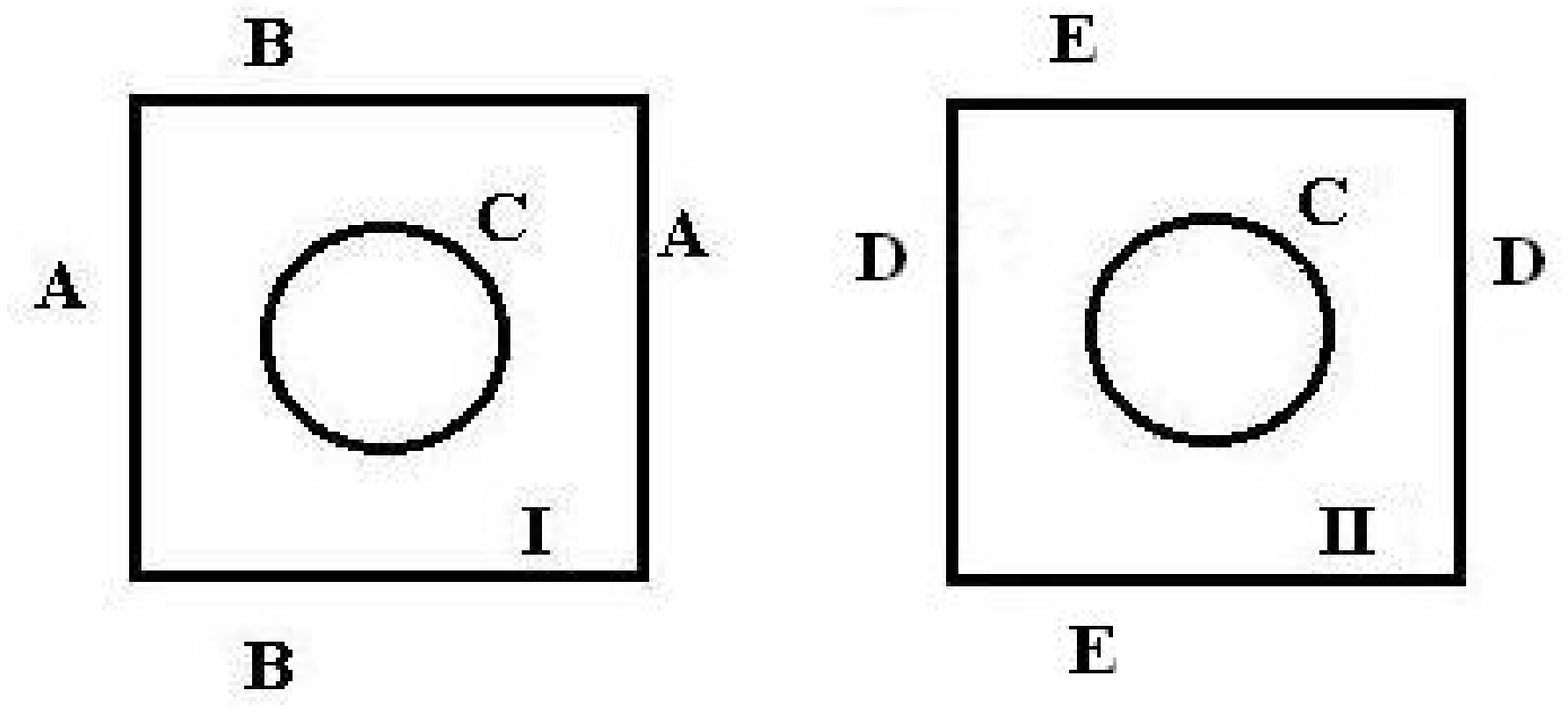}\\
  \caption{The decomposition of the double torus as two
  copies of ${\mathcal P}_a$, see text.}
\end{figure}

\begin{figure}
  \includegraphics[width=200pt]{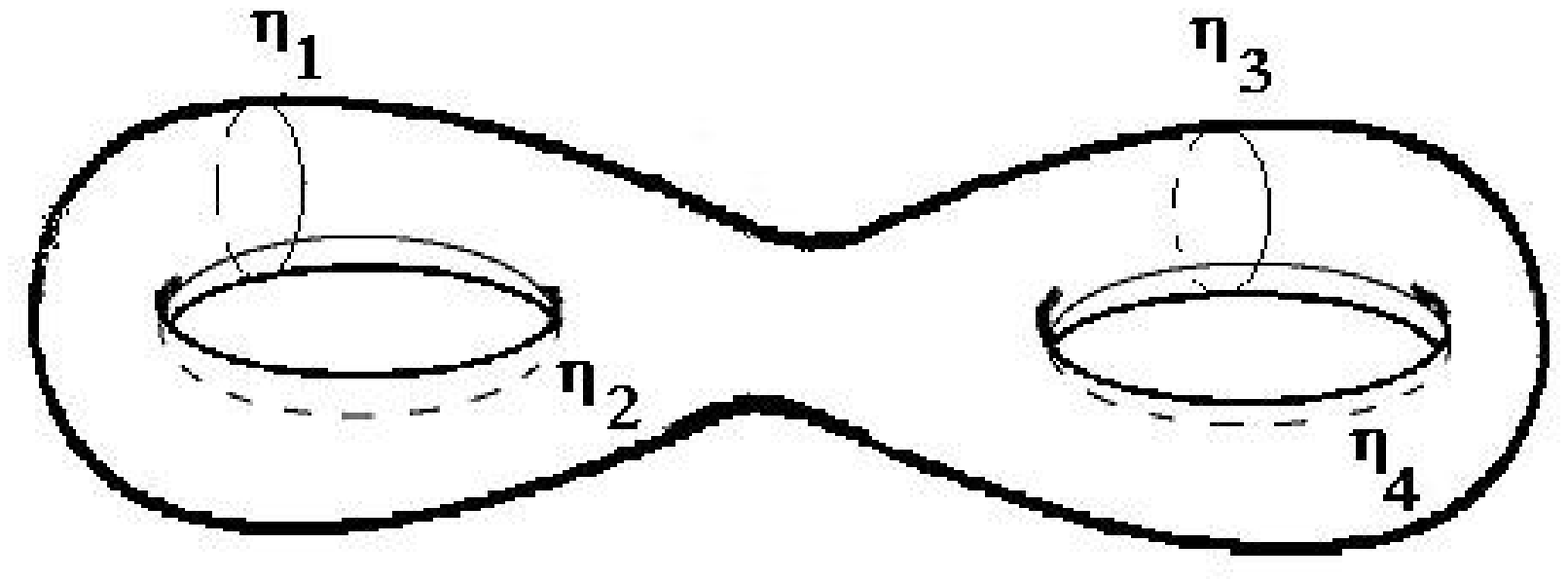}\ \
  \includegraphics[width=150pt]{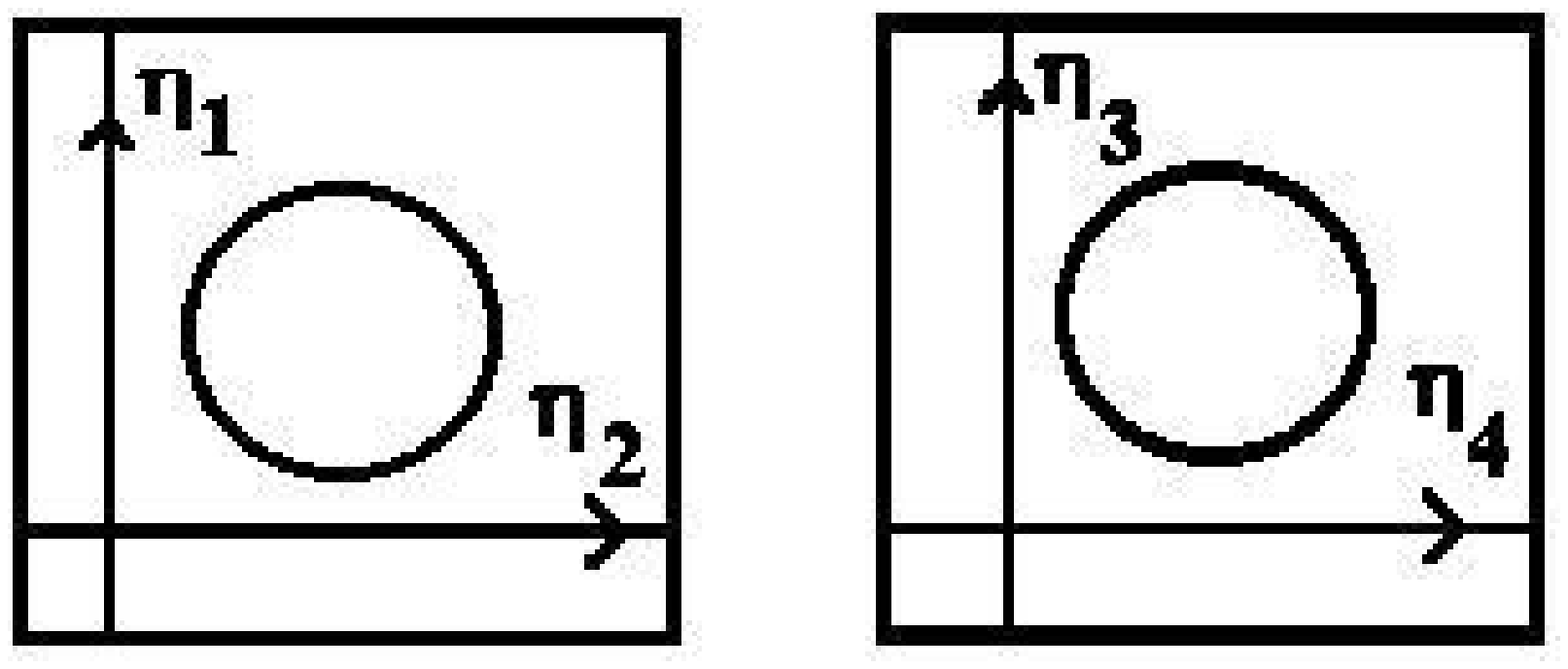}\\
  \caption{The four cycles providing a basis for the homology of
    the double torus, chosen in the natural way for the framework of
    example III (see text) and represented as in figg.3 and 4.}
\end{figure}

It suffice now to consider a simple "unperturbed" system such as
$$ \cases{
{\dot r_1} = \a (r_1 , r_2 ) & \cr {\dot r_2} = \b (r_1 , r_2 ) &
\cr {\dot \vth_1} = \om_1 (r_1,r_2) & \cr {\dot \vth_2} = \om_2
(r_1,r_2) & \cr} $$ to get, with simple hypotheses on
$\a,\b,\om_1,\om_2$ -- e.g. that there are $\rho_1,\rho_2$ such
that $\a (\rho_1,\rho_2) = 0 = \b (\rho_1,\rho_2)$, and
$\om_1(\rho_1,\rho_2) \not=0$, $\om_2(\rho_1,\rho_2) \not= 0$ --
that the system admits a double torus $\La_0$ in ${\mathcal M}$ as
an invariant manifold. This corresponds to the situation discussed
in example II.

If we then smoothly perturb the system, it is natural to ask if
$\La_0$ is somehow preserved (upon smooth deformation) in the
perturbed system. Our theorem allows to answer this question.

Finally, it should be stressed that the system considered in this
example could be hamiltonian, and more specifically a hamiltonian
perturbation of a hamiltonian integrable system -- or also,
staying within the original framework of Nekhoroshev's theorem
\cite{Nek}, a partially integrable hamiltonian system.

In this case the unperturbed system would preserve all double tori
being the double covering of (the part of) an invariant torus in
${\mathcal P}_a$, and we would be in the standard case of
perturbation of an integrable or partially integrable system;
however -- as well known -- the impact conditions would cause the
system to be generically chaotic on the invariant double torus.

With our construction we are able to deal with the case of
perturbation of integrable systems with impacts, at least for what
concerns preservation upon deformation of the invariant double
tori (the construction can also be generalized to more complex
situations).

Note that in this case one should just compute monodromy on the
standard invariant torus (see point (c) above); hence in this
framework we just extended the validity of the
Poincar\'e-Lyapounov-Nekhoroshev's theorem \cite{Nek,GaeAOP} to a
class of systems with elastic impacts.

\section*{Appendix. Explicit formulas for example II}

In this appendix we provide explicit solutions to equation (6) and
similar ones for other relevant commutators in the transition
region $A_1 \cap A_2$ (similar ones hold for $A_2 \cap A_3$). We
recall that we have chosen the vector fields to be given by
formula (5).

We define
$$ {\mathcal S} \ := \ \s_1 X_1 + \s_2 X_2 + \mu_1 Y_1 + \mu_2 Y_2 $$
(it turns out we can choose e.g. $\mu_2 = 0$) and, for ease of writing,
$$ \eta(x,y) \ := \ q(x) - \sin (y) \ ; \ \chi (x,y) := h(x) \, \eta (x,y) \ . $$

We look first for solutions to
$ [X_1 , Y_2 ] \ = \ {\mathcal S}$.
With our conventions, a solution to this is provided by
$$ \s_1 = {\cos (y) \over \eta (x,y)} \ , \
\s_2 = - {1 \over  \eta (x,y)} \ , \
\mu_1 = {q(x) \over \eta (x,y)} \ , \
\mu_2 = 0 \ . $$

The equation $ [X_1 , X_2 ] = {\mathcal S}$ is satisfied with
$$ \begin{array}{rl}
\s_1 =& - \[ h(x) q' (x) \cos^2 (y) \, + \, \sin (y) \, - \, q(x) \sin^2 (y) \] \, / \, \chi (x,y) \ , \\
\s_2 =& \[ \cos (y) \, \( h(x) q'(x) \, + \,  \sin (y) \) \] \, / \, \chi (x,y) \ , \\
\mu_1 =& - \[ \sin (y) \cos (y) \, \( h(x) q' (x) \, + \, q(x) \) \] \, / \, \chi (x,y) \ , \\
\mu_2 =& 0 \ . \end{array} $$

The equation $ [Y_1 , X_2 ] = {\mathcal S}$ is satisfied with
$$ \begin{array}{rl}
\s_1 =& - \[ \cos (y) \, \( \sin (y) \( h(x) q' (x) + q(x) \) - 1\)  \] \, / \, \chi (x,y) \ , \\
\s_2 =& - \[ \cos^2 (y) \, - \, h(x) q' (x) \sin (y) \] \, / \, \chi (x,y) \ , \\
\mu_1 =& \[ q(x) \cos^2 (y) \, - \, h(x) q' (x) \sin^2 (y) \] \, / \, \chi (x,y) \ , \\
\mu_2 =& 0  \ . \end{array} $$

Finally, the equation $ [Y_1 , Y_2 ] = {\mathcal S}$ is satisfied with
$$ \s_1 = - 1 \ , \ \s_2 = \mu_1 = \mu_2 = 0 \ . $$

Note that all of these are smooth in $B_1 \cap \B_2$, and hence in
$\^A_1 \cap \^A_2$, provided $ \chi (x,y) := h(x) [q(x) - \sin
(y)]$ is nowhere zero in $B_1 \cap B_2$.

\vfill\eject

\end{document}